\documentclass{aa}  
\usepackage{graphicx}
\usepackage{txfonts}
\usepackage[figuresright]{rotating}
\begin{document}
\title{Optical and infrared properties of V1647 Orionis during the 2003-2006 outburst}
   \subtitle{I. The reflection nebula}
\author{D. Fedele \inst{1,2}, M. E. van den Ancker\inst{1}, M. G. Petr-Gotzens\inst{1}, N. Ageorges\inst{3} \and P. Rafanelli\inst{2}
\fnmsep\thanks{Based on observations collected at the European Southern Observatory, Paranal, Chile. Proposal ID: 272.C-5045, 272.C-5046, 074.C-0679, 075.C-0489, 276.C-5022}}
\offprints{D. Fedele}
\institute{European Southern Observatory, Karl Schwarzschild Strasse 2, D-85748 Garching bei M\"unchen, Germany\\
\email{dfedele@eso.org}
\and
Dipartimento di Astronomia, Universit\'a degli studi di Padova, Vicolo dell'Osservatorio 2, 35122 Padova, Italy 
\and
European Southern Observatory, Alonso de Cordova 3107, Vitacura, Casilla 19001, Santiago 19, Chile}
%
\abstract
{}
{The recent outburst of the young eruptive star \object{V1647 Orionis} has produced a spectacular appearance of a new reflection nebula 
in Orion (\object{McNeil's nebula}). We present an optical/near infrared investigation of McNeil's nebula. This analysis is aimed at 
determining the morphology, temporal evolution and nature of the nebula and its connection to the outburst.}
{We performed multi epoch $B$, $V$, $R_C$, $I_C$, $z_{gunn}$, and $K_S$ imaging of McNeil's nebula and V1647 Ori as 
well as $K_S$ imaging polarimetry. The multiband imaging allows us to reconstruct the extinction map inside the nebula. Through 
polarimetric observations we attempt to disentangle the emission from the nebula from that of the accretion disk around V1647 Ori.
We also attempt to resolve the small spatial scale structure of the illuminating source.
}
{The energy distribution and temporal evolution of McNeil's nebula mimic that of the illuminating source. The extinction map 
reveals a region of higher extinction in the direction of V1647 Ori. Excluding foreground extionction, the optical extinction 
due to McNeil's nebula in the direction of V1647 Ori is A$_V$ $\sim$ 6.5 mag. The polarimetric measurement shows a 
compact high polarization emission around V1647 Ori. The percentage of $K_S$ band linear polarization goes from 10 -- 20 \%. The
vectors are all well aligned with a position angle of 90$\degr$ $\pm$ 9$\degr$ East of North. This may correspond to the orientation of a
possible accretion disk around V1647 Ori. These findings suggest that the appearance of McNeil's nebula is due to 
reflection of light by pre-existing material in the surroundings of V1647 Ori. We also report on the discovery of a new 
candidate brown dwarf or protostar in the vicinity of V1647 Ori as well as the presence of clumpy structure within \object{HH 22A}.}
{}
\authorrunning{D.Fedele et al.}
\titlerunning{Properties of McNeil's nebula}
\maketitle
\section{Introduction}
In late 2003 the young eruptive star V1647 Orionis in the L1630 cloud in Orion went into outburst. In the following 4 months the star 
brightened up to 6 magnitudes in the $I_{C}$ band. The rapid and strong brightness rise resembles that of a FU Orionis object. In these 
pre-main-sequence stars the outburst is thought to be triggered by a disk instability which leads the stellar accretion rate to increase
by more than four orders of magnitude on a time scale of months (e.g. Kenyon \& Hartmann \cite{kh}). The outburst of V1647 Ori produced 
the appearance of a reflection nebula known as McNeil's nebula (McNeil \cite{mcneil}). The nebula is also visible in archival data
during a previous outburst of V1647 Ori in 1966-1967 (Aspin et al. \cite{aspin}). The shape of the nebula in the optical is cometary with
V1647 Ori at its apex. In the near infrared McNeils's nebula has an arc-like morphology. This is a common feature of nebulae associated 
with FU Orionis-like objects. In these objects, it is usually thought that the nebular emission is produced by scattering in the lobe of
a bipolar structure (Goodrich \cite{goodrich}). In the case of McNeil's nebula only the lobe oriented toward us is seen while the 
opposite lobe is probably obscured by circumstellar dust. Reipurth \& Aspin (\cite{reipurth}) suggested that the bipolar structure might
be produced by powerful outflow activity from the central star. The outflow opens a cavity in its surroundings which is illuminated by 
the central star at the occurrence of an outburst. An alternative scenario is that the elongated morphology is due to the relative motion
of the star trough the parent cloud (Goodrich \cite{goodrich}). In this case there is only a single cavity, as often seen. 
\noindent 

We present here an analysis of optical/near infrared imaging and near infrared imaging polarimetry of McNeil's nebula and discuss the 
nature of this structure. 
\section{Observations and data reduction}\label{sec:obs}
We observed V1647 Ori and McNeil's nebula in the optical and near infrared using FORS2 and NACO at ESO's Very Large Telescope in Paranal,
Chile. FORS2 (\cite{appenzeller}) is an optical facility (3000-10000 \AA) equipped with two 2k4k MIT CCDs. It allows imaging in different
bands and grism spectroscopy. NACO (\cite{rousset}; \cite{lenzen}) provides adaptive optics assisted imaging, imaging polarimetry, 
coronography and spectroscopy, in the  1 -- 5 $\mu$m range. NACO's infrared camera (CONICA) is equipped with an Aladdin 1024$\times$1024 
pixel InSb array detector. 
\begin{table*}
\caption{Log of FORS2 photometric observations. The image quality refers to the seeing measured on unresolved point-sources in the 
images. Multiple short (5 -- 10 sec) and long (30 sec) exposures were taken applying an offset along the North-South direction. }
\label{table:photlog}
\centering
\centering
\begin{tabular}{ll|ll|ll|ll|ll|ll}
\hline\hline
           &            & \multicolumn{10}{c}{Exposure time [sec] \&  Image Quality [\arcsec]}\\
Date (UT)  & MJD        & \multicolumn{2}{c}{B}        & \multicolumn{2}{c}{V} & \multicolumn{2}{c}{R$_C$} & \multicolumn{2}{c}{I$_C$} & \multicolumn{2}{c}{z$_{Gunn}$} \\
\hline
2004-02-17 & 53052.082  & 10,3 x 30 & 0.91 & 10,3 x 30 & 1.18 &  5,3 x 30 & 0.85 &  5,3 x 30 & 0.65 &  5, --     & 0.67\\ 
2004-12-20 & 53359.258  & 10,5 x 30 & 0.75 & 10,5 x 30 & 0.60 &  5,5 x 30 & 0.50 &  5,5 x 30 & 0.48 &  5, --     & 0.53\\
2006-01-02 & 53737.253  & 10,5 x 60 & 0.73 & --        & --   & 10,5 x 60 & 0.58 & 10,5 x 60 & 0.50 & 10, 5 x 60 & 0.50\\
\hline
\hline
\end{tabular}
\end{table*}
\subsection{Optical photometry}
Photometry of V1647 Ori and McNeil's nebula in five different optical filters (B, V, R$_C$, I$_C$ and z$_{Gunn}$) was performed with 
FORS2 on 2004 February 17, December 20 and 2006 January 02 (no V-band photometry was done on this night). Photometric conditions were 
registered during the three nights. Multiple short (5 -- 10 sec) and long (30 sec) exposures were taken dithering the telescope in the 
North-South direction with an amplitude of 30\arcsec. A log of the observations is reported in table \ref{table:photlog}. We performed a 
standard data reduction process (bias subtraction, correction for flatfield and cosmic rays) using IRAF. The magnitude was computed using
differential aperture photometry adopting an aperture radius of 2\farcs52 (10 pixels) centered at the stellar position. The sky has been 
computed within an annulus with inner and outer radius respectively of 3\farcs02 and 4\farcs28 and subtracted from the aperture 
photometry. The computation of the stellar flux and subtraction of the local background are affected by the presence of McNeils's 
nebula. This leads to slightly different magnitude estimates of V1647 Ori depending on the telescope and instrument used. In particular, 
given the better spatial resolution of our data compared with previous data, we may better disentangle the contribution from V1647 Ori 
from that of the nebula.

\noindent

To convert the FORS2 instrumental magnitudes to the standard photometric system, photometry of standard star fields (\object{SA98} and
\object{PG0231}) was performed in the B, V, R$_C$ and I$_C$ bands during the three nights. The zero point level was measured using the 
standard stars field. The first order color dependence and atmospheric extinctions coefficients were assumed equal to the mean FORS2 
values provided by ESO\footnote{{\it http://www.eso.org/observing/dfo/quality/FORS2/}}. For the z$_{Gunn}$ filter, spectrophotometric 
standard stars were observed during the three nights: S83 in \object{[CS62] E5} (Stetson \cite{stetson}), \object{Feige 110} and 
\object{LTT 3864}. For Feige 110 and LTT 3864 the calibrated z$_{Gunn}$ magnitude was computed convolving their tabulated flux 
distribution\footnote{{\it http://www.eso.org/instruments/fors/tools/}} with the response of the filter+CCD 
combination\footnote{{\it http://www.eso.org/instruments/fors/inst/Filters/}}. For the standard star S83 in [CS62] E5, we computed the 
z$_{Gunn}$ magnitude from the empirical correlation between z$_{Gunn}$ and the VRI bands (Smith et al. \cite{smith}). We 
find a z$_{Gunn}$ magnitude of 15.49 for the star E5-S83, 12.81 for Feige 110 and 12.09 for LTT3864. The aperture photometry of 
V1647 Ori is reported in Table~2.
\noindent

The surface brightness of McNeil's nebula has been computed in two different apertures centered respectively on RA=05:46:13.051 
DEC=--00:05:56.25 (aperture radius = 7\farcs56) and on RA=05:46:14.025 DEC=--00:05:31.81 (aperture radius = 6\farcs3). These two 
locations refer respectively to blob B and C defined by Brice\~no et al. (2004). The sky background has been evaluated far from the 
nebula and subtracted from the surface brightness of the two structures. The result is reported in table.~2.
\begin{figure}
\centering
\includegraphics[scale=0.45]{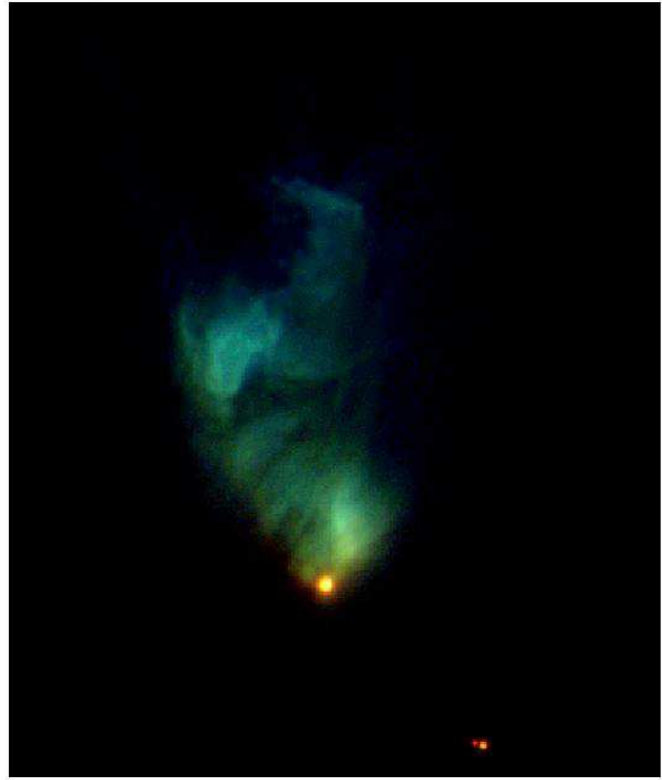}
\caption{Color image of V1647 Ori and McNeil's nebula obtained with FORS2 on 2004 December 30. The blue, green and red colors correspond 
respectively to the photometric bands B, R$_C$ and z$_{Gunn}$. The dimension of the image is 2\farcm07 $\times$ 1\farcm88. North is up, East 
is left.}\label{fig:color}
\end{figure}
\subsection{Near infrared observations}
$K_S$-band imaging was obtained with NACO on 2005 April 08. Short exposures (DIT = 0.375 s, NDIT = 80) were acquired with the S27 camera 
(pixel scale 27.15 mas/pixel, FOV 28\arcsec $\times$ 28 \arcsec) dithering the telescope at different position. The total exposure
time is 18 minutes. Thanks to the infrared wavefront sensor of NACO, the observation was adaptive optics assisted. A Strehl ratio of 0.18
was achieved resulting in an angular resolution of 0\farcs13. Observation of a photometric standard star (HD60778) was also taken. The 
reduction was carried out with the ESO pipeline V. 1.3.5\footnote{{\it ftp://ftp.eso.org/pub/dfs/pipelines/naco/}}. The raw frames were 
corrected for bad pixels and flatfield. The sky has been subtracted from the dithered images which were then shifted and co-added to 
obtain a final mosaic. 
\noindent
 
Polarimetric observations were performed with NACO on on March 01 2006 in $K_S$-band with the S27 camera. 10 multiple dithered exposures 
(DIT = 25 s, NDIT = 2) were taken with a four wire grid analyzer at four different angles (0$^{\circ}$, 45$^{\circ}$, 90$^{\circ}$, 
135$^{\circ}$). After calibration (bad pixels removal, flat fielding and sky subtraction) the dithered frames were shifted and added. The
calibrated images were combined to create the Stokes parameters images ($I$,$Q$ and $U$) and the linear polarization module ($P$) and 
position angle ($\theta$). The error associated with the polarization values in McNeil's nebula purely from shot-noise considerations
are of the order of 10 - 30 \% for both degree of polarization and position angle. The measured polarization of McNeil's nebula was 
calibrated using the standard star HD 38563C (Whittet et al. \cite{whittet}), for which the degree of polarization is known. Knowing the 
intrinsic polarization of HD 38563C we measured the degree of instrumental polarization and the zero point of position angle adopting an 
aperture of 1\farcs08 diameter. 

\section{Results}\label{sec:results}
\subsection{Optical imaging}
Figure \ref{fig:color} shows a color composite image of McNeil's nebula and of V1647 Ori taken with FORS2 on 2004 December 20. Blue,
green and red colors correspond respectively to the $B$, $R_C$ and $z_{gunn}$ photometric bands. The nebula has a typical cometary shape
extending to the North with V1647 Ori at its apex. Despite the red energy distribution of the illuminating source, the nebula emits 
mainly in the optical. The emission within McNeil's nebula is not uniform in intensity or in color. There are two main ``blobs'' 
of higher emission (respectively source B and C in Brice\~no et al. \cite{briceno}, see Fig. 2): the first is close to the star extending
to North-West. It is very bright in all the three bands. The second blob is farther away from the star in direction North-East at a 
distance of $\sim$ 35\arcsec. This structure emits mainly in $B$ and $R_C$ bandpasses and is spatially coincident with the knot A of the 
Herbig-Haro object \object{HH 22}. South-West of V1647 Ori the source \object{2MASS 05461162-0006279}, which is a visual binary, is clearly visible. 
\noindent

\begin{figure*}
\centering
\includegraphics[width=3cm, height=5cm]{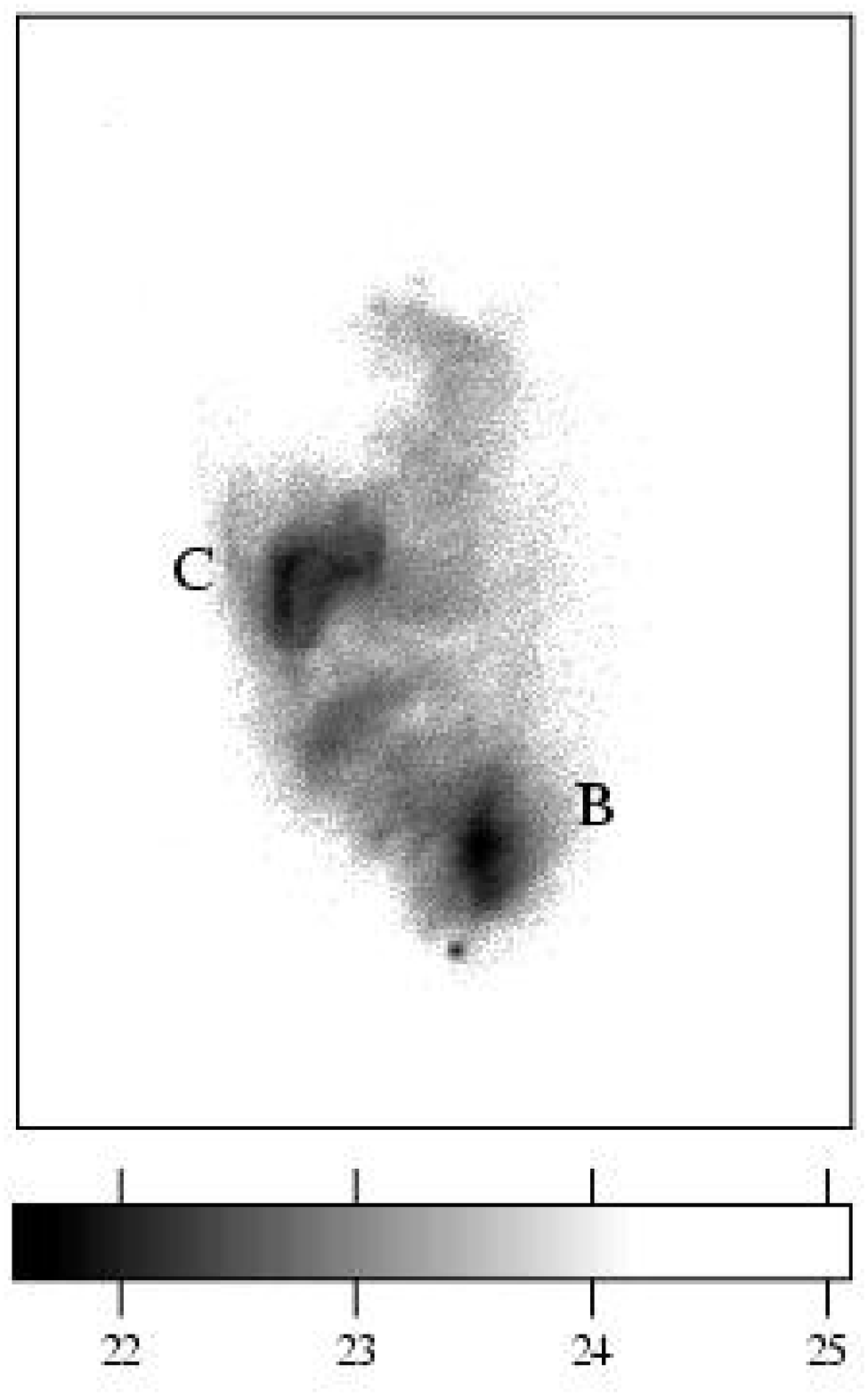}
\includegraphics[width=3cm, height=5cm]{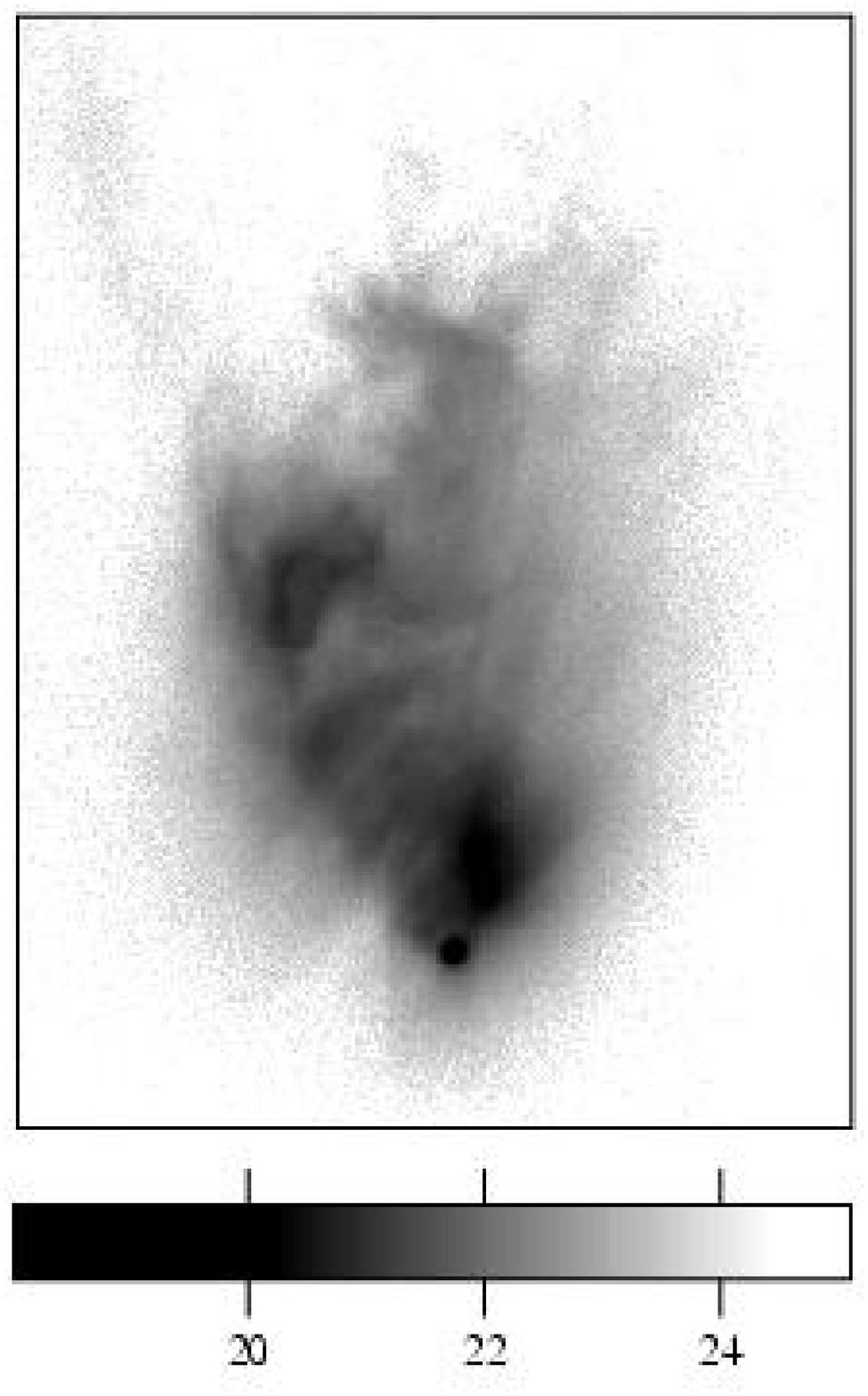}
\includegraphics[width=3cm, height=5cm]{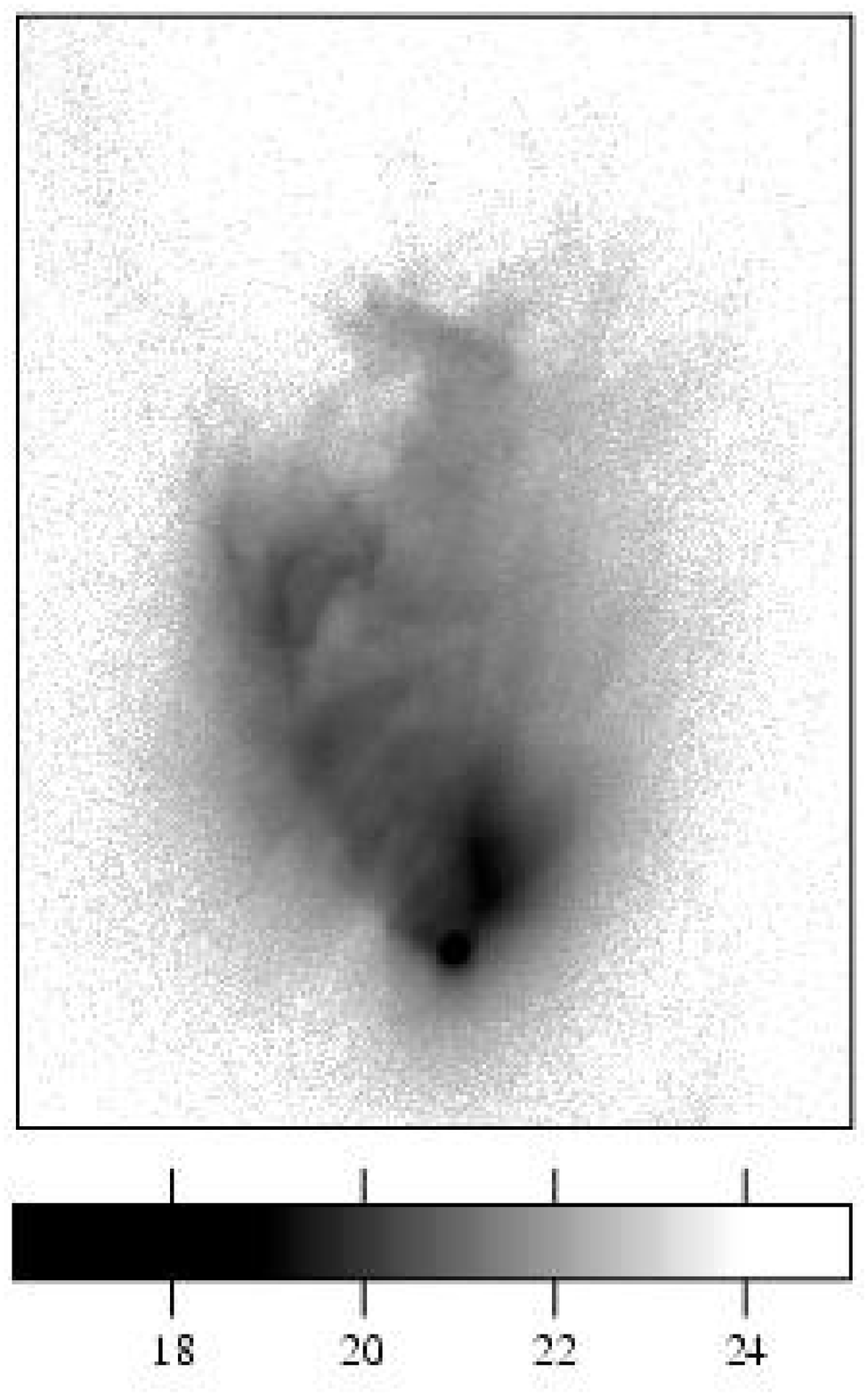}
\includegraphics[width=3cm, height=5cm]{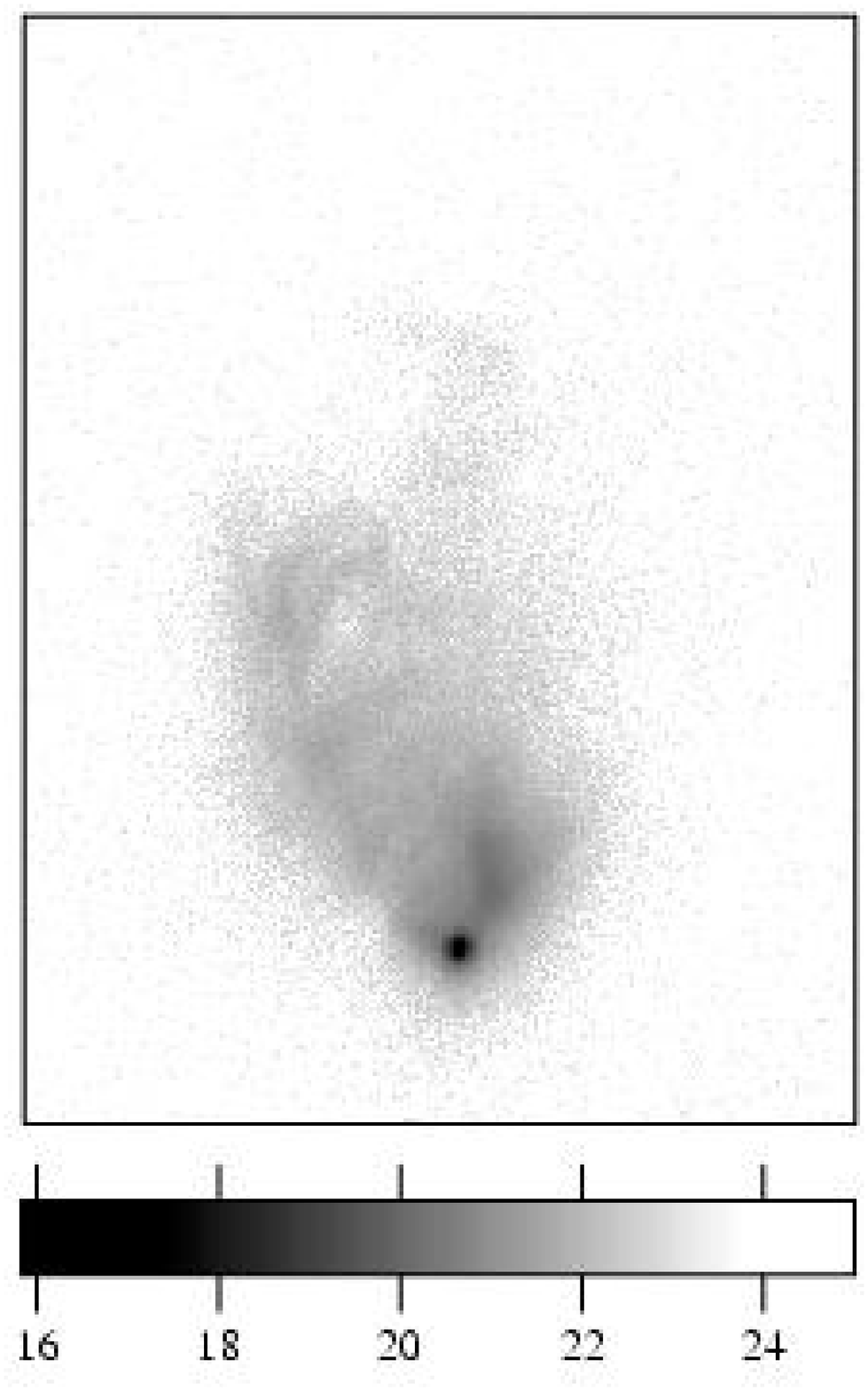}
					
\includegraphics[width=3cm, height=5cm]{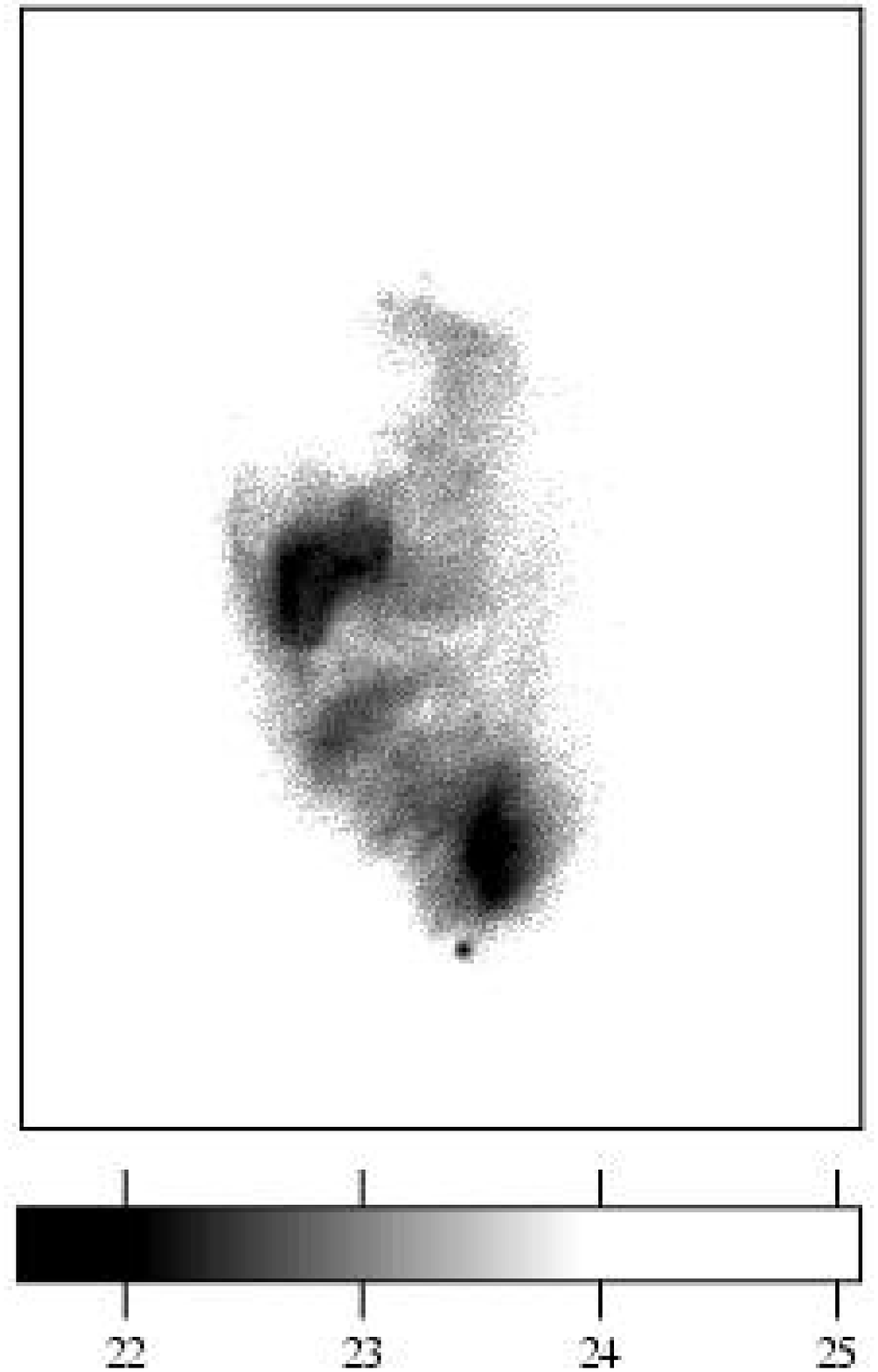}
\includegraphics[width=3cm, height=5cm]{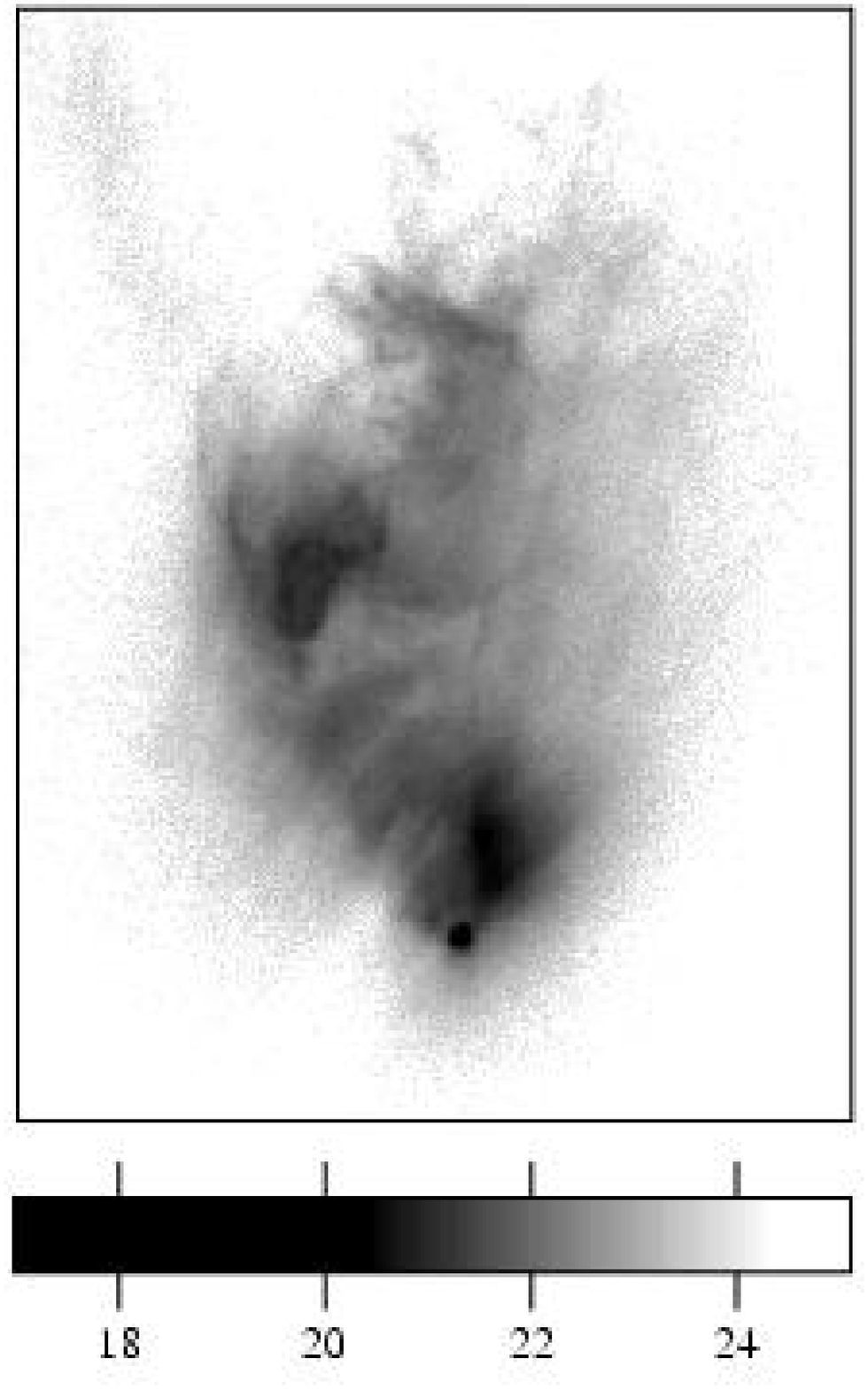}
\includegraphics[width=3cm, height=5cm]{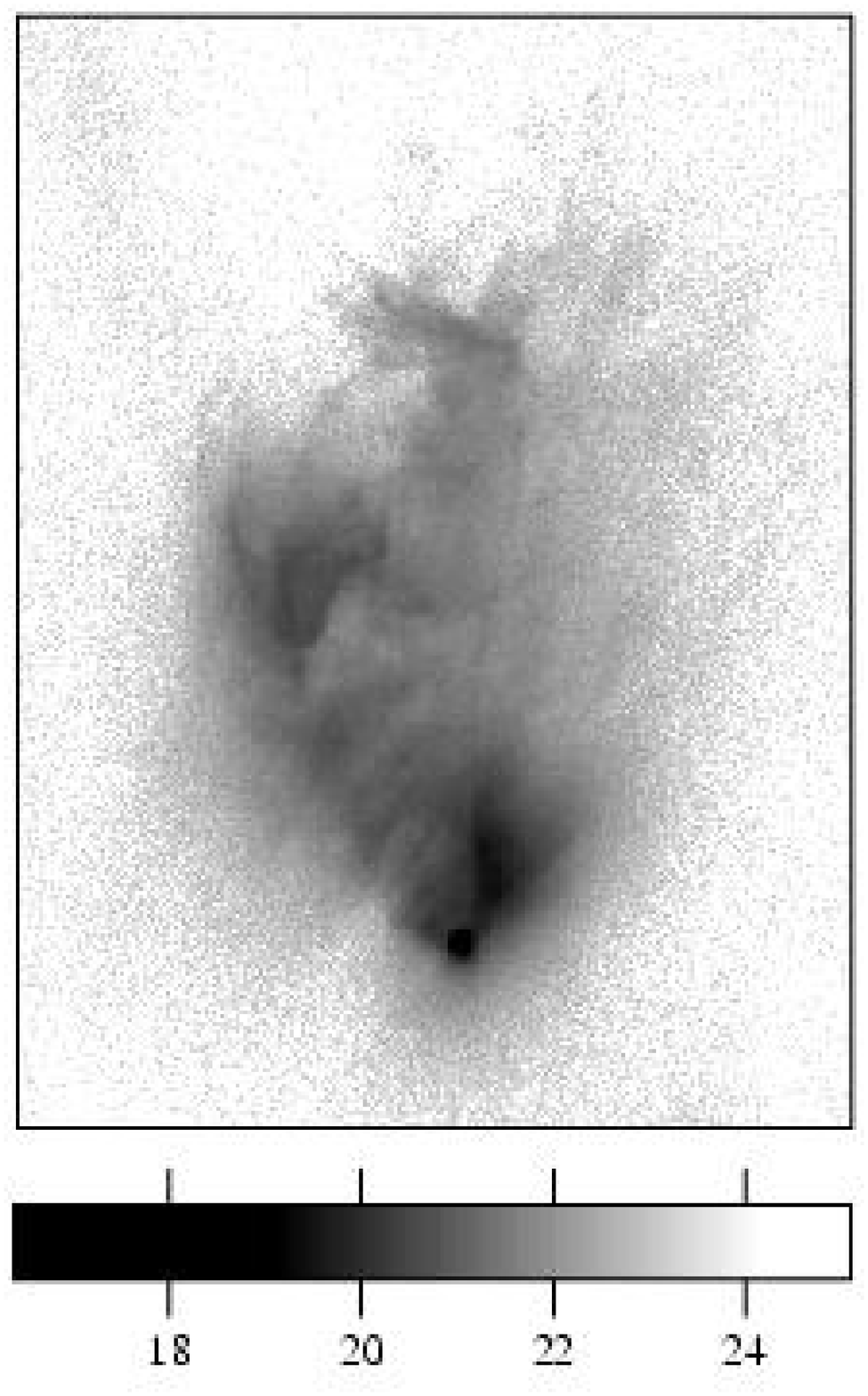}
\includegraphics[width=3cm, height=5cm]{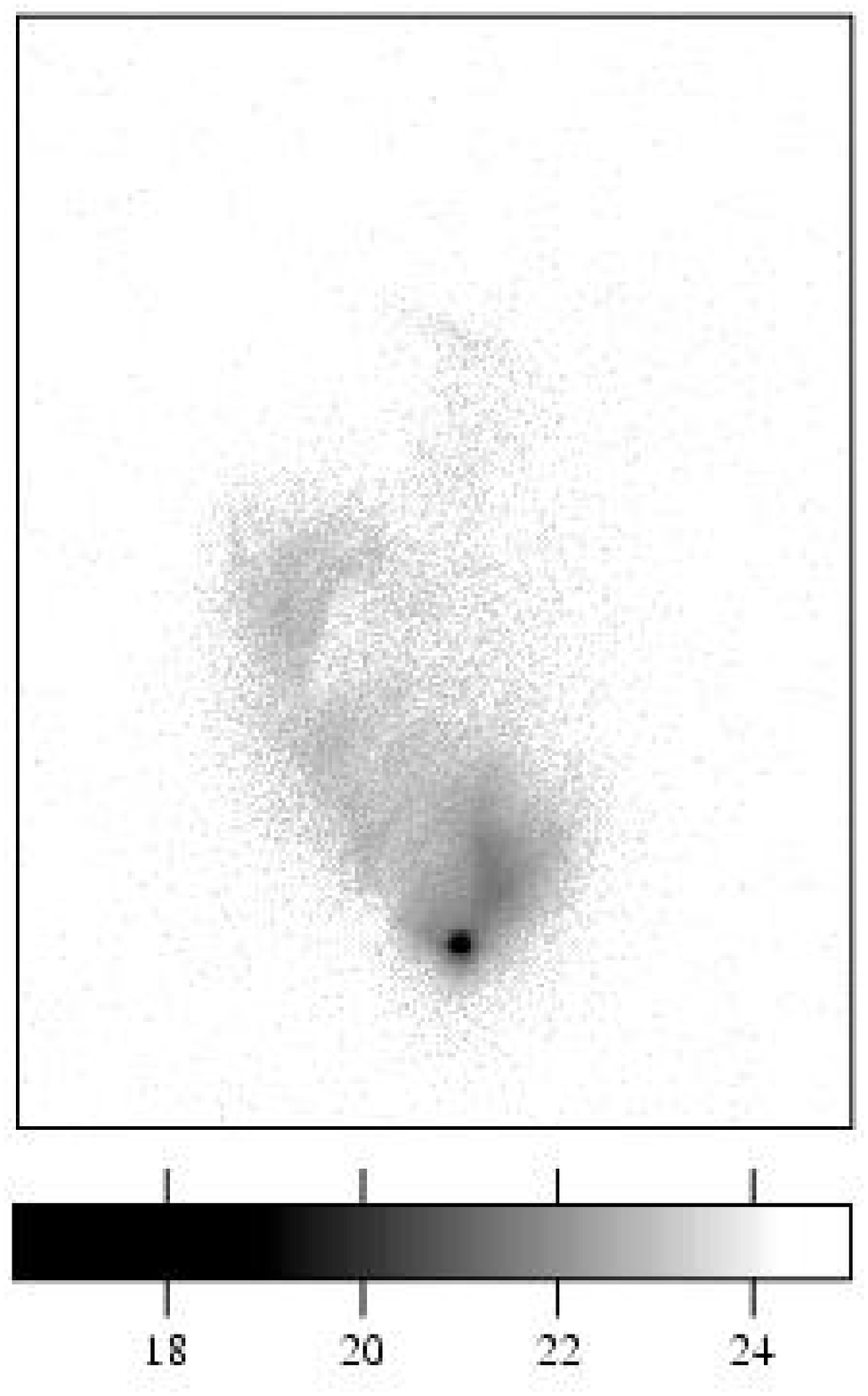}
					
\includegraphics[width=3cm, height=5cm]{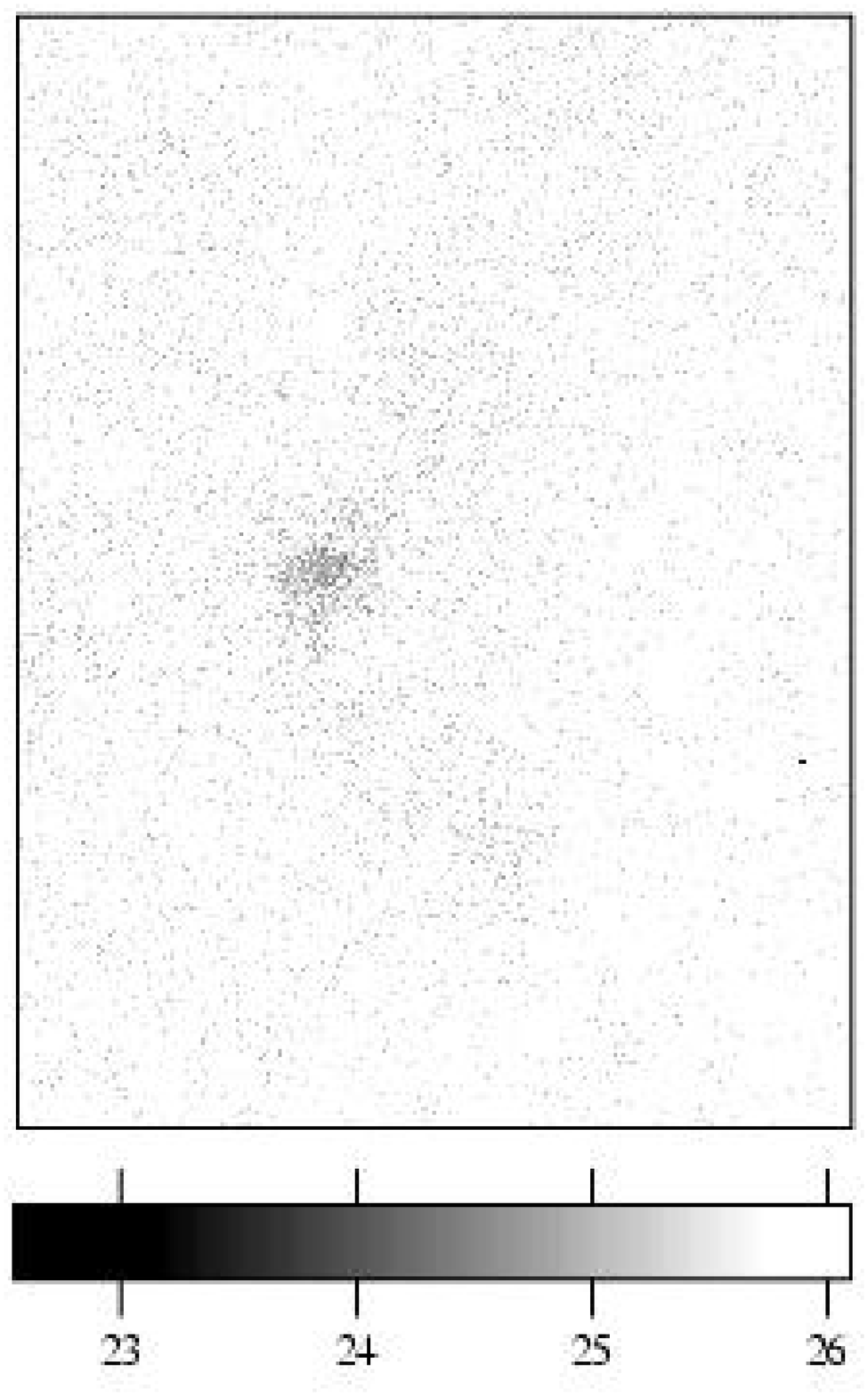}
\includegraphics[width=3cm, height=5cm]{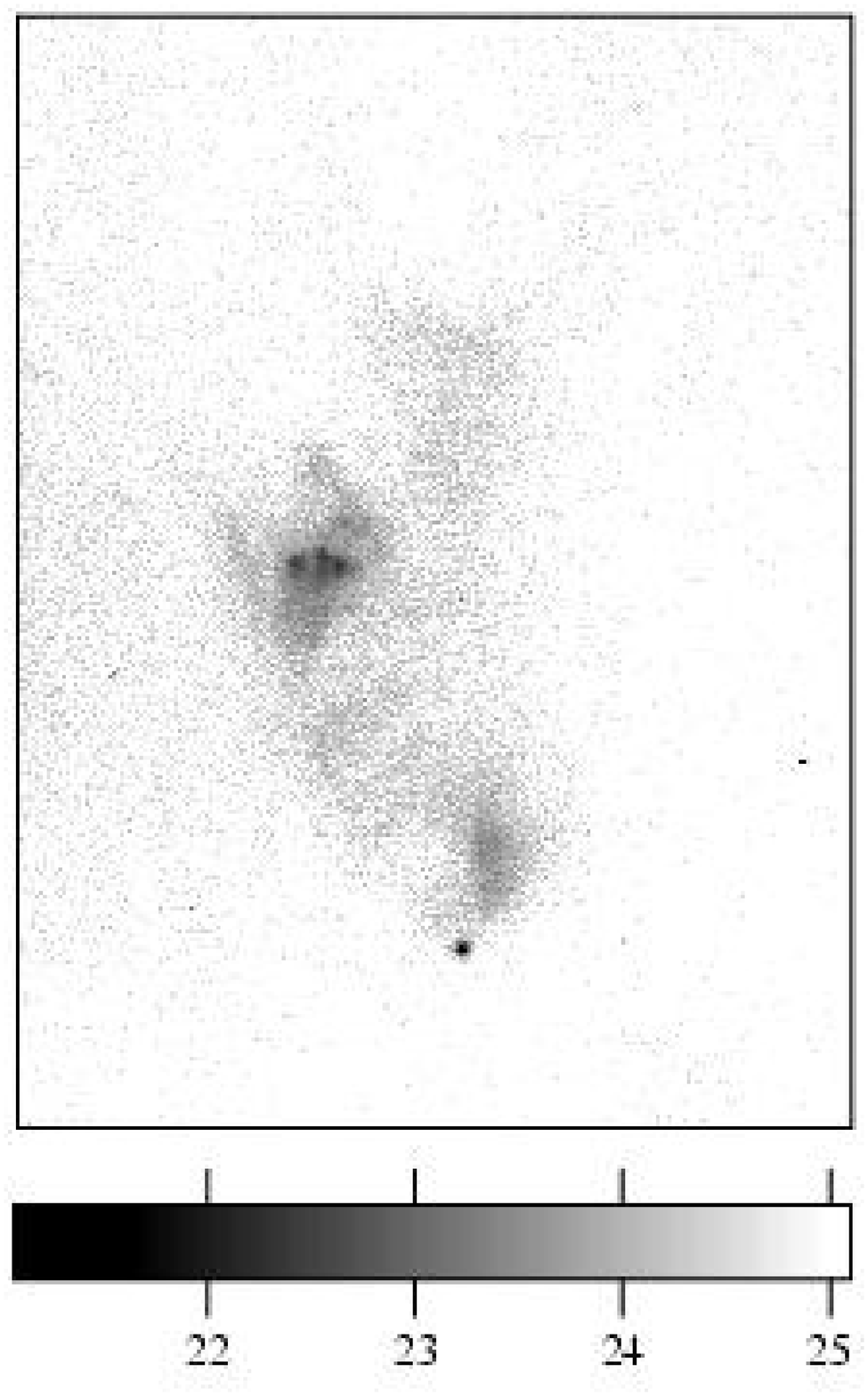}
\includegraphics[width=3cm, height=5cm]{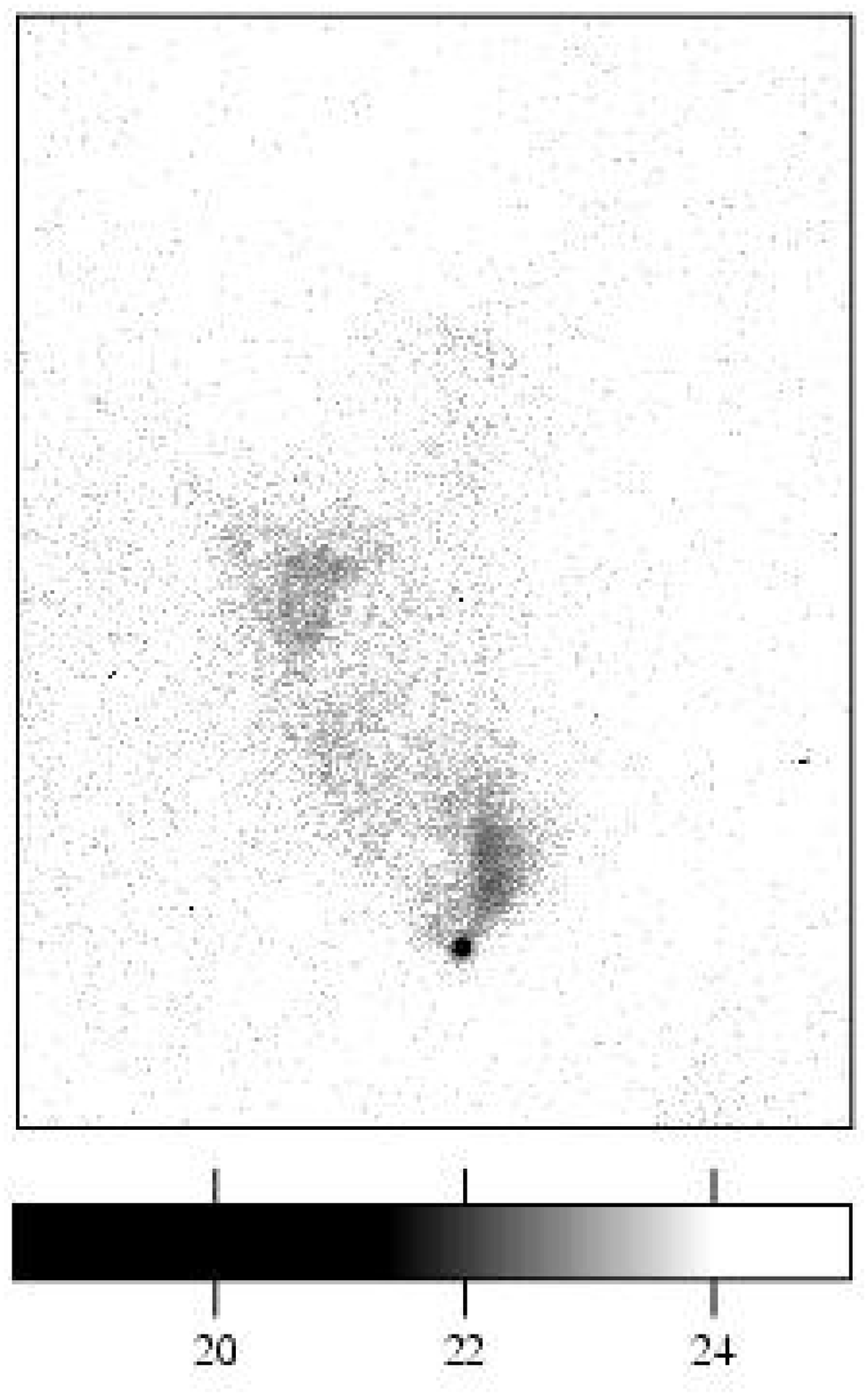}
\includegraphics[width=3cm, height=5cm]{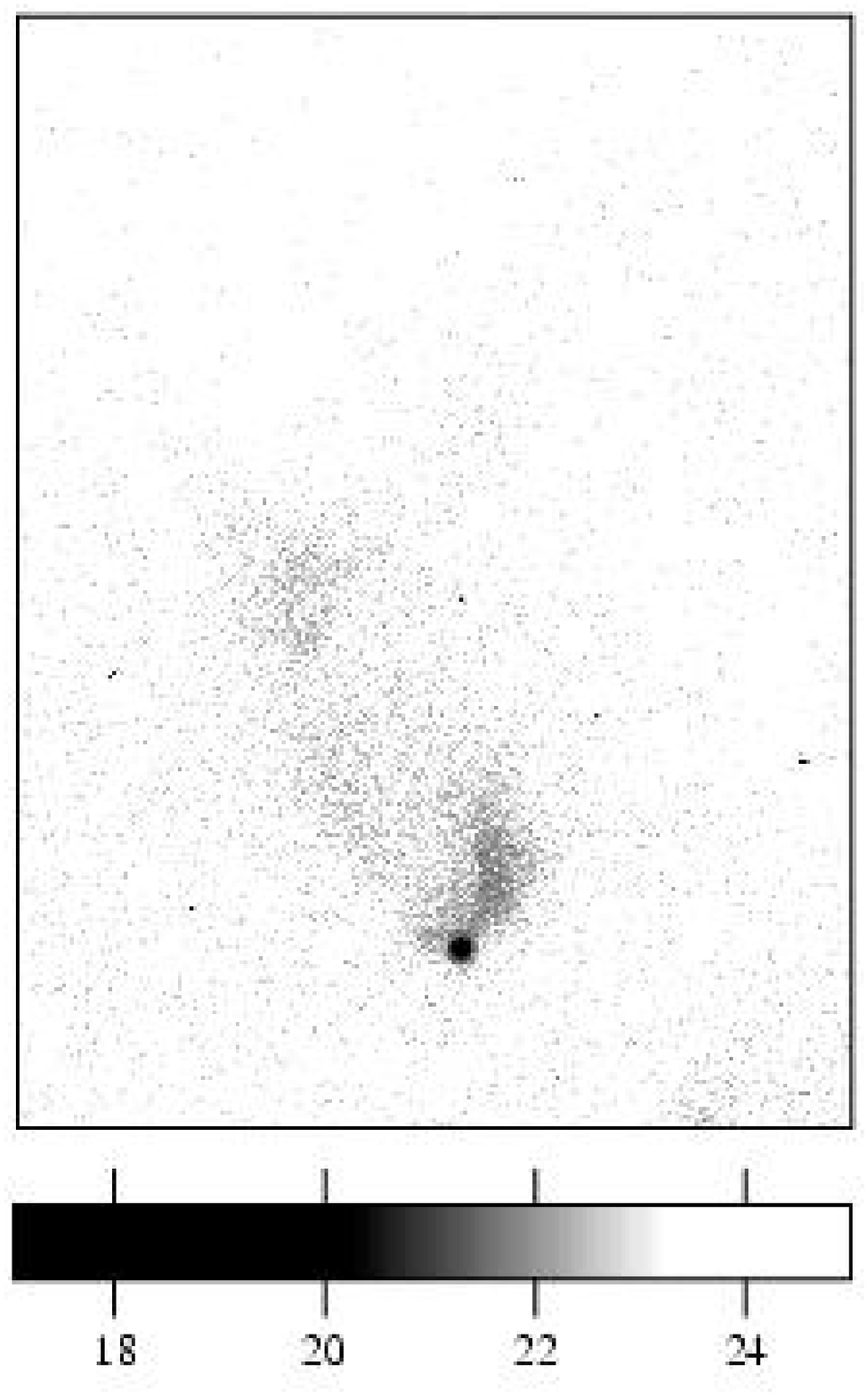}
\caption{FORS2 images of V1647 Ori and McNeil's nebula taken on 2004 February 17 (top row), 2004 December 20 (central row) and 2006 
January 02 (bottom row). The four columns corresponds to the photometric bands (from the left): B, R$_C$, I$_C$ and z$_{gunn}$. The 
images are expressed in mag/arcsec$^{2}$. North is up, East is left. All images have the same dimension of 76\arcsec x 101\arcsec.}
\label{fig:imaging}
\end{figure*}

\begin{figure}[!ht]
\centering
\includegraphics[angle=90, width=9cm]{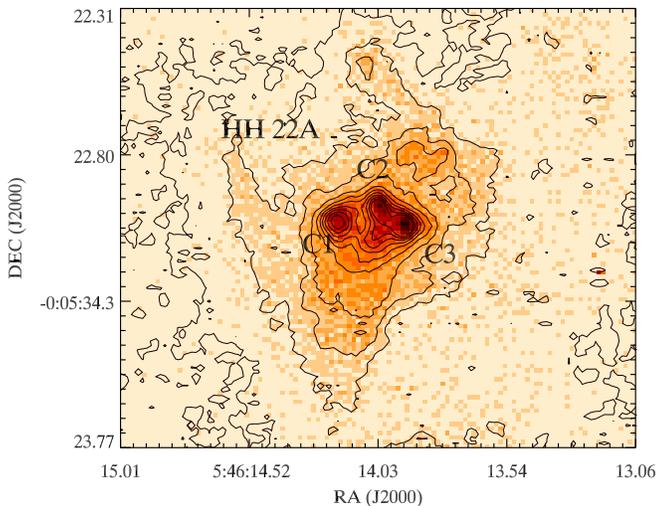}
\caption{Blow-up of the $R_C$-band image taken on 2006 January 01 showing the presence of three clumps (labeled C1,C2 and C3) in the 
vicinity of HH 22A.}
\label{fig:hh22}
\end{figure}

In Figure \ref{fig:imaging} we show a temporal sequence of images of V1647 Ori and McNeil's nebula in 4 photometric band-passes: $B$, 
$R_C$, $I_C$ and $z_{gunn}$, taken on 2004 February 17, December 20 and 2006 January 02. The pixel values were converted in surface
brightness and are expressed in mag/arcsec$^{2}$. During the first two epochs V1647 Ori was at the maximum light of the outburst 
({\em plateau} phase). On January 2006 the star was quickly fading returning at the quiescent phase. The brightness temporal evolution of
MCNeil's nebula follows that of the outbursting star: the nebular emission remains unchanged during the {\em plateau} phase, as is clear 
from the top and middle rows of Figure \ref{fig:imaging}. In early 2006 the nebula has mostly disappeared. It is no longer visible in 
$B$, where also V1647 Ori was not detected up to a limiting magnitude of $B$ = 24.9. A faint emission from blob B and C is still visible 
in the $R_C$, $I_C$ and $z_{Gunn}$ filters. The $R_C$ bandpass reveals a clumpy structure of the northern blob never seen before. The 
brightness level is of the same order as that of V1647 Ori. Given the spatial coincidence with HH 22A, such emission is likely produced 
by H$\alpha$ and forbidden lines (all falling in the $R_C$ bandpass) within the Herbig Haro knot. Figure \ref{fig:hh22} shows a blow-up 
of the $R_C$ band image taken on January 2006. Apart from a diffuse emission, three main clumps are visible in the knot A of HH 22 
(clumps C1, C2 and C3). 
\noindent

We compared the appearance of McNeil's nebula in the period February 2004 -- January 2006. The images were spatially matched using 
four different (point-like) stars as reference objects. This gives us an accuracy of roughly half a pixel (0$\farcs$13). The overall 
morphology of the nebula (including the sub-structures B and C) does not show major changes during such period. Given the FWHM of the
FORS2 images ($<$ 0$\farcs$85 in $R_C$) and the nearly two years of time interval, we conclude that no evidence of spatial motion was 
identified within McNeil's nebula down to a resolution of 0$\farcs$43 yr$^{-1}$, corresponding to an upper limit to the projected 
expansion velocity of 800 km s$^{-1}$ at the adopted distance toward V1647 Ori of 400 pc.
\begin{table*}
\caption{FORS2 aperture photometry of V1647 Ori and surface brightness of blob B and C.}
\label{table:photres}
\centering
\begin{tabular}{llccccc}
\hline\hline
Date       & MJD       & B                 & V                 & R$_C$            & I$_C$            & z$_{Gunn}$       \\
\hline
\multicolumn{7}{c}{{\bf V1647 Ori}}                                                                                     \\
           &           & (mag)             & (mag)             & (mag)            & (mag)            & (mag)            \\
\hline
2004-02-17 & 53052.075 & 20.66 $\pm$ 0.10  & 18.80 $\pm$ 0.05  & 16.90 $\pm$ 0.05 & 15.07 $\pm$ 0.03 & 13.90 $\pm$ 0.10 \\ 
2004-12-20 & 53359.248 & 20.75 $\pm$ 0.12  & 18.82 $\pm$ 0.05  & 17.06 $\pm$ 0.03 & 15.11 $\pm$ 0.04 & 13.95 $\pm$ 0.10 \\
2006-01-02 & 53737.245 & $>$ 24.9          & --                & 21.33 $\pm$ 0.13 & 18.95 $\pm$ 0.08 & 16.29 $\pm$ 0.10 \\
\hline
\multicolumn{7}{c}{{\bf Blob B}}                                                                                        \\
          &            &(mag/arcsec$^2$)   &(mag/arcsec$^2$)   &(mag/arcsec$^2$)  &(mag/arcsec$^2$)  &(mag/arcsec$^2$)  \\
\hline
2004-02-17 & 53052.075 & 22.45 $\pm$ 0.05  & 21.45 $\pm$ 0.03  & 20.69 $\pm$ 0.03 & 19.71 $\pm$ 0.02 & 20.96 $\pm$ 0.10 \\
2004-12-20 & 53359.248 & 22.90 $\pm$ 0.06  & 21.87 $\pm$ 0.03  & 21.05 $\pm$ 0.03 & 20.11 $\pm$ 0.02 & 22.19 $\pm$ 0.10 \\
2006-01-02 & 53737.245 & $>$24.9           & --                & 24.21 $\pm$ 0.13 & 23.20 $\pm$ 0.10 & 22.53 $\pm$ 0.10 \\
\hline
\multicolumn{7}{c}{{\bf Blob C}}                                                                                        \\
          &            &(mag/arcsec$^2$)   &(mag/arcsec$^2$)   &(mag/arcsec$^2$)  &(mag/arcsec$^2$)  &(mag/arcsec$^2$)  \\
\hline
2004-02-17 & 53052.075 & 22.59 $\pm$ 0.04  & 21.91 $\pm$ 0.03  & 21.41 $\pm$ 0.03 & 20.82 $\pm$ 0.03 & 22.37 $\pm$ 0.10 \\ 
2004-12-20 & 53359.248 & 22.83 $\pm$ 0.06  & 22.08 $\pm$ 0.03  & 21.52 $\pm$ 0.03 & 20.99 $\pm$ 0.03 & 23.35 $\pm$ 0.10 \\ 
2006-01-02 & 53737.245 & $>$24.9           & --                & 23.86 $\pm$ 0.10 & 23.49 $\pm$ 0.10 & 23.29 $\pm$ 0.10 \\ 
\hline
\hline
\end{tabular}
\end{table*}
\subsection{NACO imaging}
\begin{figure}[!ht]
\centering
\includegraphics[width=8.5cm]{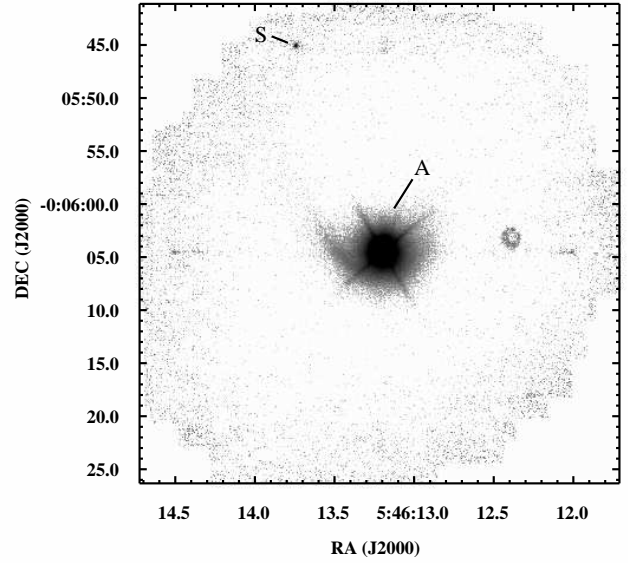}
\caption{NACO $K_S$ band image of V1647 Ori and McNeil's nebula taken on 2005 April 04. The nebula (A) appears much more compact 
than in the optical with a slightly flattened morphology in direction North-South. An arc-like structure is also detected. A point-like 
source (S, RA = 05:46:13.75, DEC = -00:05:44.5) is detected 21\farcs2 North-North East of the nebula. The circular spot West 
of V1647 Ori is a ghost image caused by the brightness of V1647 Ori.}
\label{fig:k}
\end{figure}

\begin{figure}
\centering
\includegraphics[angle=90, width=9cm]{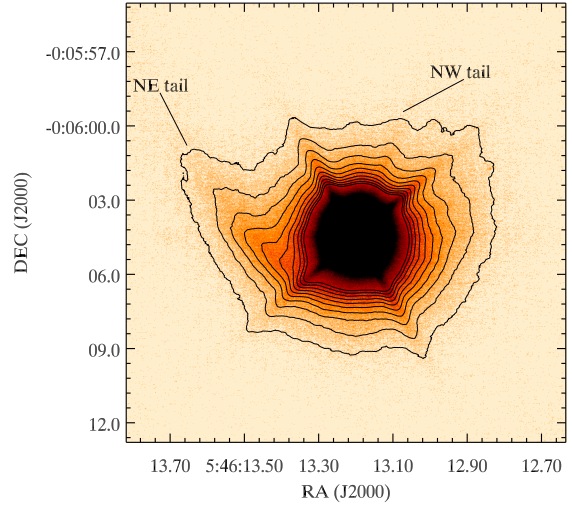}
\caption{Blow-up of the $K_S$ band image of V1647 Ori and McNeil's nebula taken on 2005 April 04. The two tails characterizing the 
ark-like structure are showed. On spatial scales smaller than 1\farcs5 no deviation from circular profile is detected.}
\label{fig:k_zoom}
\end{figure}

Figure \ref{fig:k} shows the $K_S$-band image of V1647 Ori and McNeil's nebula taken on 2005 April 08. Compared to the optical image, 
the nebula in the $K_S$ band appears much more compact. It has a slightly flattened morphology in the North-South direction, being more 
elongated in the direction perpendicular to the optical emission. An arc-like tail is visible from the East side of the compact nebulosity
extending in direction North-East. Figure \ref{fig:k_zoom} is a blow-up of the small spatial scale of McNeil's nebula. Apart from the 
diffraction patterns, another tail extends from the central star in direction North-West. This is fainter than the NE one and less 
extended. Compared to previous observations at equal wavelength (Reipurth \& Aspin \cite{reipurth}; Ojha et al. \cite{ojha05}, 
\cite{ojha06}; Acosta-Pulido et al. \cite{acosta}), all taken between February and November 2004, the morphology of McNeil's nebula is 
overall unchanged. However, in our NACO image the nebula is fainter and less extended. Despite the better angular resolution of this 
observation compared to previous work we could not find evidence of small spatial scale structure within the central 1\farcs5, or 600 AU
assuming a distance V1647 Ori of 400 pc. The magnitude of the system (star+nebula) within an aperture diameter of 16\farcs3 is 
$K_S$ = 8.97 $\pm$ 0.01. 
\noindent

A point-source 21\farcs2 North of the nebula is clearly visible in figure \ref{fig:k} (RA(J2000) = 05:46:13.75, DEC(J2000) = -00:05:44.5).
We will discuss the nature of this source in Appendix A.

\subsection{NACO polarimetry}
In our $K_S$-band polarimetric observation (Fig.~\ref{fig:polmap}) we detect a compact polarized emission centered on V1647 Ori. We 
do not detect polarized emission at radii $>$ 1\farcs6 from V1647 Ori. The images were binned in 2$\times$2 square bins. Alignment and 
subtraction residuals are present in the inner region as well as along the diffraction pattern of the nebula. The polarization values go 
from 10 -- 20 \%. The vectors are well aligned with a mean position angle of 90$\degr$ $\pm$ 9$\degr$  East of North. The highest values 
are detected North and South of V1647 Ori. The most likely reason for not detecting extended polarized light from McNeil's nebula could
be that we did not integrate deep enough in the four exposures. Acosta-Pulido et al. (2007) have indeed measured large scale polarized
emission from the nebula in their LIRIS/WHT $J$ band observations. 
\section{Data interpretation}
Following the method of Magnier et al. (\cite{magnier}, hereafter MA99), we can use the color dependence of scattered light to probe the
distribution of material inside the nebula and in the vicinity of V1647 Ori. As in MA99, our basic assumption is that the light from 
V1647 Ori is coming directly to the earth from the central star without being scattered but only partially extinguished. On the other 
hand the light from the nebula that reaches the earth is also scattered. The effect of scattering would be that the observed colour
of V1647 Ori is bluer than the colour of V1647 Ori seen by the nebula. As a consequence, the nebula would appear redder than V1647 Ori. 
Since there are no parts of McNeil's nebula which are redder than V1647 Ori we believe that our assumption is a good one. We have:
\begin{equation}\label{equ:neb1}
f_*(\lambda) = f_{*,0}(\lambda) 10^{-0.4A_{V,1}(\frac{A_{\lambda}}{A_{V}})}
\end{equation} 
\begin{equation}\label{equ:neb2}
f_{neb}(\lambda) = f_{*,0}(\lambda) 10^{-0.4A_{V,2}(\frac{A_{\lambda}}{A_{V}})} \Gamma \big( \frac{\lambda}{\lambda_0} \big) ^{-\gamma}
\end{equation} 
where $f_*$ and $f_{neb}$ are the observed flux of V1647 Ori and of the nebula, $\Gamma$ is a normalization factor which takes into 
account the amount of scattered light and $\gamma$ the wavelength dependence of the scattering process. For Rayleigh scattering 
$\gamma$ =  4. Simulations of other types of scattering processes (such as Thomson or Mie) produce values close to 4 (MA99). From 
the ratio of the two observed fluxes we have:
\begin{equation}\label{equ:neb3}
log \Big( \frac{f_{neb}(\lambda)}{f_*(\lambda)} \Big) = +0.4 \Delta A_V \frac{A_{\lambda}}{A_V} + log \Gamma - \gamma log \big( \frac{\lambda}{\lambda_0} \big)
\end{equation} 
Equation \ref{equ:neb3} can be solved to determine $\Delta A_V$, i.e. the difference between the extinction towards the line of 
sight of V1647 Ori ($A_{V,1}$) and the extinction towards any direction of McNeil's nebula ($A_{V,2}$). All images were smoothed and 
binned in order to have the same PSF and to increase the S/N in the nebula (final pixel scale 0\farcs5 / pixel). Counts were
converted in flux units (Jy) using the appropriate magnitude-flux relation. The standard extinction law, E(B-V) = 3.1 A$_V$ has been 
applied to compute $A_{\lambda}$. Finally we performed a linear least-squares in the five images on a pixel by pixel basis to determine
$\Delta A_V$. Figure \ref{fig:ext} shows the result. The background image is the V image of McNeil's nebula. The contour plot is 
the differential extinction. The extinction is not uniform in the nebula. Close to V1647 Ori and at the base of the nebula 
$\Delta A_V$ is lower. Moving from the star to the North-East, a region of higher extinction shows up. In correspondence of the northern
knot, the method used here to compute the extinction may not be valid as it is coincident with knot A of the Herbig-Haro object HH 22 
which is intrinsically bright. The total optical extinction in the direction of V1647 Ori caused by material within McNeil's nebula 
is $\sim$ 6.5 mag. As this estimate does not include foreground extinction, it is a lower limit to the total optical extinction towards 
V1647 Ori. 
\noindent

The differential extinction map shows some similarities with the K-band image of McNeil's nebula (Fig. \ref{fig:k}). The contours of 
$\Delta A_V$ in the vicinity of V1647 Ori show an arc-like geometry. Moreover, the region of higher extinction in the nebula mimics the 
tails seen in the NACO image. The opening angle between the two tails of higher extinction is $\simeq$ 95$\degr$. 
The asymmetry in the spatial distribution of dust in McNeil's nebula may be indicative of an outflow activity with the material moving
away from the star towards the North. 
\noindent

The result does not change if we apply the Cardelli et al. (\cite{cardelli}) extinction law with larger values of R$_V$ up to 5. This is 
not surprising beacuse changing the value of $R_V$ has a significant effect only on the shape of the ultraviolet extiction and produces
only small changes in the optical/NIR.
\noindent

\begin{figure}
\centering
\includegraphics[width=8cm]{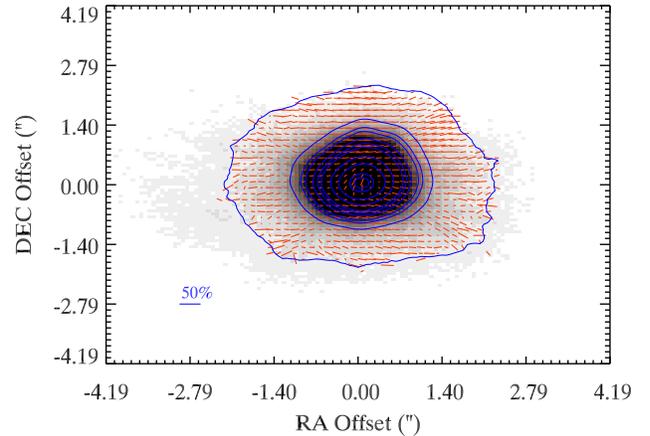}
\caption{$K_S$ band imaging polarimetry of V1647 Ori and McNeil's nebula taken on 2006 March 01. Polarization vectors are superimposed
upon total intensity map and contours. Alignment and subtraction residuals are present the inner region as well as along the diffraction
pattern of the telescope. The polarization values go from 10 -- 20 \%. The highest values are detected North of V1647 Ori. The mean 
position angle of the vectors is 90$\degr$ $\pm$ 9$\degr$. North is up, East is left.}
\label{fig:polmap}
\end{figure}

\begin{figure}
\centering
\includegraphics[angle=90, width=10cm]{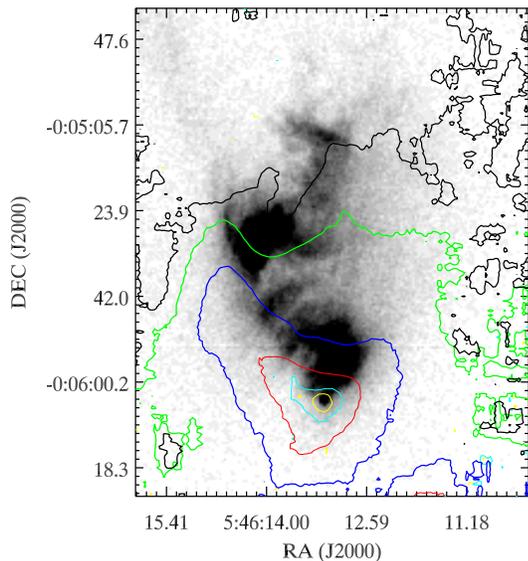}
\caption{Differential extinction map of McNeil's nebula. The grayscale shows the $V$-band image of the nebula. Overplotted are the 
$\Delta A_V$ contour level 1.5 (yellow), 2.5 (cyan), 3.5 (red), 4.5 (blue), 5.5 (green), 6 (black) magnitudes. The extinction increases
gradually as the line of sight from the star to the nebula tilts towards the East.}\label{fig:ext}
\end{figure}

Our $K$ band polarization map of V1647 Ori reveals a compact region of aligned vectors with high degree of polarization. At larger scales 
the polarization pattern is centro-symmetric as found by Acosta-Pulido et al. (2007). Such structures are often seen in near-infrared 
polarimetric map of Class I YSOs with circumstellar nebulae. These systems show a region of aligned vectors, known as a ``polarization
disk'', at the location of the central source, and a gradual transition to a centrosymmetric pattern of vectors in the surrounding 
nebula. The polarization disc is often attributed to multiple scattering in cases where the optical depth toward the central source is 
too high for direct observation (e.g Whitney \& Hartmann. 1993; Bastien \& Menard 1988). Multiple scattering, however, is not able
to reproduce high polarization ($>$ 15 \%) measured sometimes in low optical depth regions (Lucas et al. \cite{lucas} and references 
therein). The aligned vectors of the polarization disk are usually parallel to the disk plane. Interestingly, the position angle of the 
observed polarization in the vicinity of V1647 Ori (90\degr $\pm$ 9\degr, Fig.~\ref{fig:polmap}) is perpendicular to the major axis of
the reflection nebula seen in the optical. If the large scale reflection nebula can be interpreted as being shaped by a wind or outflow 
from the central star, the polarization vectors would indeed be aligned with the disk plane. 
\noindent

Moreover, the high percentage of linear polarization observed toward V1647 Ori may suggest that the central (proto-)star is obscured even
at 2.2 $\mu m$.

\smallskip
\noindent
\section{Discussion and conclusions}
Young stars are often surrounded by optical or infrared nebulae (e.g. Padgett et al. \cite{padgett}, Zinnecker et al. \cite{zinnecker}). 
Given the complexity of the star formation process there are many mechanisms to produce such nebulae and their nature is not unique. 
Strong outflow activity from the central star is responsible for the typical Herbig Haro objects which emit mainly in forbidden lines 
as the shocked gas cools. Strong UV flux from a massive star ionize all the circumstellar environment producing an \ion{H}{II} region.
Also wind from a young star might produce large scale nebular emission. In the case of McNeil's nebula, the energy distribution and the 
temporal evolution of the nebula mimics that of the illuminating source. The polarization pattern and extinction map do agree with dust 
scattering taking place in a circumstellar environment.
\noindent

All these findings suggest that McNeil's nebula is a pure-reflection nebula where the light from the central star and the accreting disk 
is scattered by pre-existing material in the vicinity of V1647 Ori. 
\noindent

The morphology of McNeil's nebula resembles that of FU Orionis-like objects, which often show an arc-like morphology (Goodrich 
\cite{goodrich}). The prevalent theory about the nature of these objects (Goodrich \cite{goodrich}) is that the nebular emission mimics 
the lobe of a bipolar structure in which one of the lobe is oriented toward the observer. The secondary lobe may be obscured by the 
circumstellar disk and/or envelope. The nebula will always have an arc-like morphology unless it is observed pole-on. In this case the 
nebula will appear ring-like. Reipurth \& Aspin (\cite{reipurth}) suggested that the bipolar structure might be produced by powerful 
outflow activity from the central star. The outflow opens a cavity in its surroundings which is illuminated by the central star at the 
occurrence of an outburst. The arc-like structure seen in the K-band image and its similarity with the extinction map suggest
that an outflow travels roughly from the star to the North (slightly twisted to North-West). The current asymmetry in the distribution of
material within McNeil's nebula could be the result of the clearing of pre-existing material (perhaps left over from the original cloud)
by a previous outflow. 
\noindent 

It is worth to note that from an optical spectrum of V1647 Ori taken on January 2006 (when the star was quickly returning to its 
pre-outburst flux level), Fedele et al. (\cite{fedele}) found strong emission from forbidden lines which may indeed indicate ongoing 
outflow activity. From our multi-epoch imaging of McNeil's nebula we derive an upper limit to the motion of material within McNeil's 
nebula of 800 km s$^{-1}$.

\begin{appendix}
\section{On the nature of the red source North-East of V1647 Ori.}
The point-like source 21\farcs2 North North-East of V1647 Ori (RA = 05:46:13.75, DEC = -00:05:44.5, Fig. ~\ref{fig:k}) is not 
visible in any of our optical images up to a $R_C$ limiting magnitude of  $\simeq$ 24.5. From the NACO observation we compute a magnitude
of $K_S$ = 16.63 $\pm$ 0.13 within an aperture diameter of 1\farcs08. Acosta-Pulido et al. (\cite{acosta}) detected the source only in 
the $K$ band, but not in the $J$ band and only barely in the $H$ band. They estimate magnitudes of $J > 19$ and $H > 16$. The same object
is clearly visible in the SPITZER/IRAC 4.5 $\mu$m image of Muzerolle et al. (\cite{muzerolle}), although they don't give an estimate
of its magnitude.
\noindent

An unreddened stellar photosphere of any temperature has a $J - K$ color of $<$ 1 mag (e.g. Kenyon \& Hartmann \cite{kh95}). We derive 
$J - K > 2.37$. To redden a colour of $J - K = 1$ to the observed lower limit an extinction of $A_K = 0.5$, or equivalently $A_V = 6.9$,
is needed. Thus, if we assume that this object is located in the background of the Orion cloud, its absolute K-band magnitude would
be $<$ 16.1 mag. This excludes the possibility that it is a main sequence star. It may be an highly reddened background giant star, 
although the space density of these stars makes the a priori chance of finding such an object low.  
\noindent

The very red energy distribution of this source ($R_C - K_S > 7.9$) is also compatible with either a brown dwarf (for which $R - K >$ 
6 mag, Kirkpatrick et al. \cite{kirkpatrick}) or a protostar still embedded in its infalling envelope. Assuming that it is located in 
Orion, just behind McNeil's nebula, we can estimate its foreground optical extinction from the extinction map of figure \ref{fig:ext}.
The position of this source is within the $\Delta A_V = 4$ contour level, i.e. the extinction due to McNeil's nebula toward this 
direction is A$_V$ $\simeq$ 1.5 mag. Assuming a further contribution of A$_V$ = 1.5 mag due to the foreground interstellar medium, the 
total optical extinction is A$_V$ $\simeq$ 3 mag (or A$_R$ $\simeq$ 2.3 mag). The corrected energy distribution is still very red 
($R_C - K_S >$ 5.7 mag) and both the brown dwarf and protostar hypothesis are still valid. 
\noindent

Whatever its nature, embedded protostar, brown dwarf or highly reddened background giant, this source located close to V1647 Ori is an 
unusual object which deserves further study.

\end{appendix}
\begin{acknowledgements}
The authors thank the ESO staff for performing the service mode observations. We are grateful to an anonymous referee for many
useful comments and suggestions. 
\end{acknowledgements}

%

\end{document}